\documentclass[journal]{IEEEtran}
\usepackage{cite}

\ifCLASSINFOpdf
  \usepackage[pdftex]{graphicx}
\else
\fi
\usepackage{pgfplots, pgfplotstable}

\usepackage[T1]{fontenc}
\usepackage[utf8]{inputenc}
\usepackage{tikz}
\usetikzlibrary{shapes,decorations}
\usepackage{environ}

\usepackage[cmex10]{amsmath}

\usepackage{mathtools}
\usepackage{algorithm}
\usepackage{algpseudocode}

\usepackage{array}
\usepackage{blkarray}

\usepackage{mdwmath}
\usepackage{mdwtab}

\usepackage{eqparbox}

\usepackage[tight,footnotesize]{subfigure}

\usepackage[font=footnotesize]{subfig}
\usepackage{stfloats}
\usepackage{threeparttable}
\usepackage{verbatim}

\usepackage{acronym}
\usepackage{color}

\pgfplotsset{compat=newest}
\begin{document}
%
\title{{Rational Agent based Decision Algorithm for Strategic Converged Network Migration Planning}}
%
%
%

\author{Sai~Kireet~Patri,
        Elena~Grigoreva,
        Wolfgang~Kellerer,~\IEEEmembership{Senior~Member,~IEEE},
        Carmen~Mas~Machuca,~\IEEEmembership{Senior~Member,~IEEE}
\thanks{Sai Kireet Patri is with ADVA Optical Networking SE and a member of the research team at the Chair of Communications Networks, Technical University of Munich, Germany. Elena Grigoreva, Wolfgang Kellerer and Carmen Mas Machuca are with the Chair of Communications Networks, Technical University of Munich, Germany. This work has received funding by the German Research Foundation (DFG) under the grant numbers MA6529/2-1 and KE1863/4-1. }}

%
%

\markboth{Journal of Optical Communications and Networking,~Vol.~XX, No.~XX, XXXX}%
{Patri \MakeLowercase{\textit{et al.}}: Rational Agent based Decision Algorithm for Strategic Converged Network Migration Planning}
%



\maketitle

\begin{abstract}
To keep up with constantly growing user demands for services with higher quality and bandwidth requirements, telecommunication operators are forced to upgrade their networks. This upgrade, or migration of the network to a new technology, is a complex strategic network planning problem that involves techno-economic evaluations over multiple periods of time. The state-of-the-art approaches consider migrations to a concrete architecture and do not take uncertainties, such as user churn, into account. This results in migration cost underestimations and profitability over-estimations. {In this paper, we propose a generic migration algorithm derived from a search based rational agent decision process that can deal with uncertainties and provides the migration path using a maximized utility function.} The algorithm maximizes the {migration} project profitability, measured as accumulated Net Present Value~(NPV). This {flexible and generic} methodology has been evaluated on the example of migration from existing copper networks to the future-proof Passive Optical Network~(PON) architectures. {Our proposed flexible migration algorithm is validated over pure residential and converged scenarios in a fully reproducible case study. The results yeld that the migration flexibility is a key to the profit maximization.} 
\end{abstract}


%
\IEEEpeerreviewmaketitle

\section{Introduction}\label{sec:introduction}
%
%
%
%
\IEEEPARstart{B}{andwidth}-hungry services are becoming a daily reality for every household. Online gaming, HD-TV and an overall growth of employees working remotely from their homes challenge the network operator with ever growing bandwidth requirements. The operators are forced to upgrade the network by the competition and governmental initiatives \cite{FTTHCouncilEurope2015}, while fighting to maintain a  sufficient gap between costs and the revenues. This upgrade, or migration, to the next technology while satisfying user and regulatory requirements calls for careful strategic network planning to fulfill the requirements, while maximizing the benefits of the operator~\cite{mcmahon1999strategic}. In general, strategic network planning is done as a telecommunication project evaluation, which allows choosing the suitable technology, evaluating the risks and planning the upgrades of the network, i.e., migrations. 

{A migration project (from here on, referred to simply as "project")} consists of a multi-dimensional, multi-period planning problem and its solution involves market penetration forecasting, dimensioning of the network infrastructure and processes, as well as evaluation of their Total Cost of Ownership~(TCO) \cite{Reyes2014a}. Due to migration planning complexity and lack of open network TCO models, the state-of-the-art migration models are case study specific, focusing either on only a single migration technology or a single migration path. Throughout this work, the term migration path refers to the technology sequence (in time) from the starting technology to the goal of the migration.
\begin{figure}[htbp!]
  \begin{center}
    \centering
    \captionsetup{justification=centering}
    \includegraphics[width=\columnwidth]{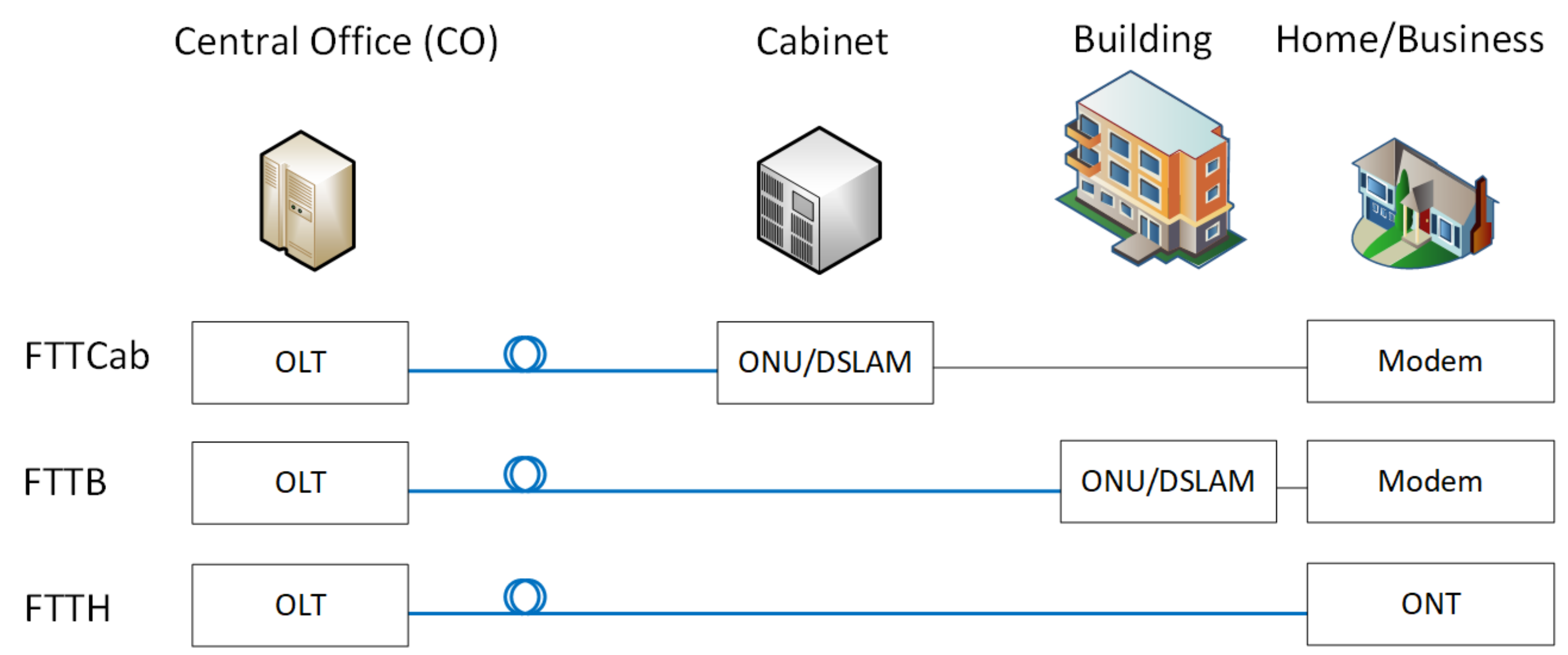}
    \caption{Fiber-To-The-x (FTTx) schematics where $x$ stands for the optical fiber termination point, which could be a cabinet~(FTTCab), building~(FTTB) or home (FTTH).}
    \label{fig:fiber_dep}
  \end{center}
\end{figure}

{Over the years, tree based Artificial Intelligence~(AI) algorithms have emerged as a front-runner to solve problems in which complete information is not available. In the case of access network migration, this maps to providers taking strategic decisions as to when and how many migration steps are needed in order to maximize the profits, keeping in mind an uncertain number of subscribers, variations on revenues and also associated TCO. The solution proposed in this work is based on the Expectimax Search algorithm~\cite{Klein2010}, which is commonly used in AI-based applications and for the first time it has been applied to the network migration problem, to the best of our knowledge.}

{These algorithms usually consist of (i) agents, which execute an action based on the environment, and (ii) an environment which changes based on agent's actions. Any agent which takes an action in order to maximize its performance metric, is said to be a \textbf{Rational Agent} \cite{Russell2010}.}


{The contribution of this paper is as follows. First, we propose a generic Rational Agent based algorithm to maximize the migration project profitability while taking into account user uncertainty. The introduced algorithm takes migration decisions (when to migrate to which technology) keeping in mind the migration goal, which in our case is to maximize the project profitability over the network lifetime. This algorithm aims to serve as a strategic analysis tool to help network planners, researchers and industry managers to make decisions, compare solutions, estimate time and costs, evaluate impact of new solution, etc. The proposed algorithm is validated with realistic case studies: migrating from copper based networks to Passive Optical Networks (PONs) in different deployment areas (rural, urban and dense urban) and user mixes (residential and converged planning). The validation includes creating a database of PON technologies and architectures~\cite{patri2018}, calculating their realistic costs based on geographical data (using the network planning tool available at~\cite{elenagit2018}) and finally, performing a comprehensive sensitivity analysis to validate the assumptions}.

Our case studies results show that the network operators can achieve higher profitability while providing the required data-rates to different subscribers with the flexible migration end state. In this case, the final state is chosen by the migration algorithm to maximize economic value instead of targeting a concrete technology. Furthermore, accounting for user churn allows avoiding overestimating the project profitability. In this paper, we present selected results for the urban area, the full set of results can be found in~\cite{patri2018}. {Last but not least, an important contribution of this paper is the full reproducibility of the study, as all the implementations are publicly available~\cite{elenagit2018, saigit2018} as well as a complete database of the techno-economic parameters~\cite{patri2018}. }

The rest of this paper is organized as follows. Section \ref{sec:rel_work} deals with the current state of the art for the various approach to migration planning in access networks. Section~\ref{sec:methodology} describes the problem formulation and assumptions for the migration algorithm. It lists and explains all the necessary inputs to conduct a migration study. Section~\ref{sec:mig_model} introduces the migration methodology based on {an uninformed search algorithm.} Section~\ref{sec:results} shows the results (expected Net Present Values (NPVs) and migration paths) of selected scenarios as well as the sensitivity analysis. Finally, Section~\ref{sec:conclusions} presents the conclusions and outlook.

\section{State of the Art}\label{sec:rel_work}

Optical Access Network technologies provide, among other benefits, longer reach and higher sustainable data-rates, which makes them an ideal candidate to provide high speed data-rates to customers \cite{VanderMerwe2009}. However, these optical access networks come in various configurations, each with their own type of equipment and dimensioning. A simple modelling of costs cannot suffice in finding the best technology to be deployed. Since deployment of optical access networks lasts over a long period of time (between five and ten years), an analysis involving time value of money is required to find out a cost-effective solution.

The authors of \cite{VanderMerwe2009} undertake a cost-benefit analysis of various optical network technologies, assumed to be deployed in a Fiber-to-the-Home~(FTTH) configuration, as shown in Fig. \ref{fig:fiber_dep}. However, the possible benefits of Fiber-to-the-Building~(FTTB) and Fiber-to-the-Cabinet~(FTTCab) deployments are not covered. Also, the lack of granularity in Operational Expenditures~(OPEX) calculations as well as the use of geometric models instead of geographical models for the network deployment, leads to unpredictable TCO under- and over-estimations~\cite{mitcsenkov2013geometric}. Another work applied to FTTH (i.e., without the possibility to choose the final migration state) is proposed by \cite{Reyes2014a}. The authors contribute to planning of multi-step migration, when multiple technologies are present. The output is then an optimal migration path, which considers migration window as well as holding time in every intermediate technology. However, the solution does not deal with the uncertainties such as user churn.

The use of AI based heuristic search in access network migration planning is studied in the work done by \cite{Turk2013} and \cite{Stefan}. The problem statement defined in that work is to optimize the migration process from VDSL to FTTC GPON in an urban access network, undertaken in a pre-defined migration period. However, no evidence is provided if this optimization suffices for multi-step migration when different migration options are present.

Another approach to identify the best network migration was undertaken by \cite{Tahon2012} using strategic analysis of a Real Options Approach \cite{Mun2006}. In the studied business case of a Fiber-To-The-x~(FTTx) migration from a full copper deployment, this translates to options like the size of cabinets to deploy, the services to be provided to subscribers, etc., where all the options form a decision tree. However, the authors focus on a specific case study and model the probability based on assumptions made about the per-cabinet take up rate, which cannot be generalized. 

From the literature, the need for a migration model which provides the flexibility of choosing different access network deployment options in various scenarios, while catering for revenue uncertainty is evident. {To tackle this, we use the Expectimax Search~\cite{Klein2010}, which is a simplified uninformed adversarial search tree used to model sequential games. This search method is different from informed or deterministic searches, like MinMax search~\cite{Klein2010}, because it considers the uncertainty arising from input parameters for which complete information is not available. In our work, we prefer an uninformed search over a deterministic search, because it allows us to model subscriber uncertainty in the future.}

Expectimax Search allows us to have a stochastic model for this uncertainty by maximizing the probabilistic average over all possible outcomes over time in order to find an optimal solution. Its tree based structure allows for three different nodes, namely maximizer nodes (which finds the maximum of its children), chance nodes (which finds the probabilistic average of its children) and terminal nodes (nodes which signify the end of tree building).  An interested reader may refer to Section 2.3.3 in \cite{patri2018} and lecture notes in \cite{Klein2010} for further in-depth discussion.

\vspace{+0.5em}

\section{Problem formulation, Assumptions and Input}\label{sec:methodology}
This section introduces the problem formulation, lists and explains the assumptions and necessary input for the model. Further, this section introduces the economic parameters used to evaluate the results.

\subsection{Problem Formulation}

For better understanding, we explain the migration problem on a realistic example: migration from copper based Asymmetric Digital Subscriber Line 2+ (ADSL2+) network to PON. As has been already established in Section \ref{sec:introduction}, migration of access networks is a multi-dimensional and complex process and gives the network operator multiple options to explore.

\begin{figure}[htbp!]
  \begin{center}
    \centering
    \captionsetup{justification=centering}
    \includegraphics[width=\columnwidth]{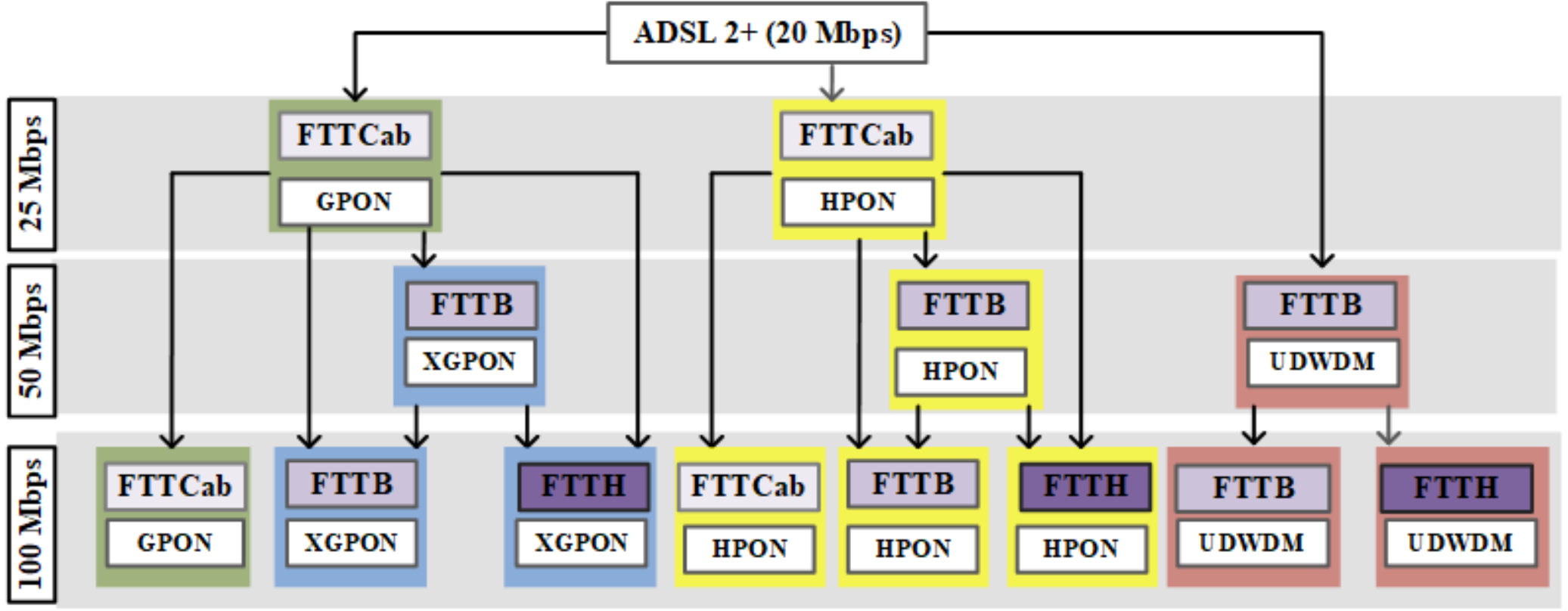}
    \caption{Schematics of the allowed migrations starting from an ADSL 2+ solution offering 20 Mbps to different solutions offering 100 Mbps, differing on the FTTx architecture, data rates and technology.}
    \label{fig:mig_tech}
    \vspace{-1em}
  \end{center}
\end{figure}

With no loss of generality, to limit the scope of the paper we concentrate on the business model of the Vertically Integrated Operator~(VIO)~\cite{Forzati2014}. {We define the VIO to be the only internet provider in a given area, who owns the physical infrastructure (fiber, ducts, remote nodes), network (OLTs, ONUs, splitters) and  as well as the services (Internet, television, telephony).} The VIO assumes to have copper based technologies already deployed in an area and looks to migrate to any PON based technology, shown in Fig. \ref{fig:mig_tech}, since these are now mature and affordable to deploy \cite{OpticalAccessSeamlessEvolution2011,VanderWee2014}. The complete migration is limited in time, which is referred as migration window \textbf{$T_{mig}$}~\cite{Reyes2014a}. However, the cost and revenue calculations take the network life-cycle into account (usually longer than $T_{mig}$), which is denoted by \textbf{$T_{NW}$}.

The business model also involves the inclusion of different types of subscriber demands, such as Residential, Business and Public Intelligent Transportation System~(ITS) LTE Macro Base Stations~(MBSs). With these, we create two different planning scenarios, namely \textit{pure residential} (i.e., connecting residential users) and \textit{converged} (i.e., connected residential, business and ITS MBS demands). In the converged planning scenario, we ensure that the ITS MBSs are assigned separate wavelengths to support their backhaul traffic for high bandwidth and security purposes.

For different PON technologies, deployments can vary based on how far the fiber drop point is from the Central Office~(CO). As shown in Fig. \ref{fig:fiber_dep}, FTTCab has Optical Network Unit~(ONU) placed at street cabinets, whereas FTTB has a fiber drop point directly inside the building. This invokes a difference in civil works cost, which as described in further sections, is an important cost driver. Another important cost driver is the cost of active equipment at the CO, using which the same PON deployment can achieve higher data-rates.

Depending on the type of technology used, Gigabit-ethernet Passive Optical Network~(GPON), 10 Gigabit-ethernet Passive Optical Network~(XGPON), Ultra Dense Wavelength Division Multiplexing Passive Optical Network~(UDWDM-PON) and Hybrid Passive Optical Network~(HPON) can be deployed~\cite{OpticalAccessSeamlessEvolution2011}. Each of these technologies have different network equipment costs.

PON architectures can offer subscribers data-rates of 25~Mbps, 50~Mbps and 100~Mbps as shown in Fig.~\ref{fig:mig_tech}. The data rate can be increased incrementally by adding more active equipment at the CO, making use of the dark fiber and by blowing additional fiber through a duct, which was assumed to be less expensive as compared to digging and closing the ducts again~\cite{OASE2013}.

For consistency, each scenario is referred as FTTx\_technology\_bitrate. For example, FTTCab\_GPON\_25 refers to the FTTCab solution using GPON and offering 25~Mbps to the end users. This way, using the migration tree shown in Fig. \ref{fig:mig_tech}, we are able to find 13 different PON deployments, for example, FTTB\_XGPON\_50, FTTB\_UDWDM\_100, etc. 

Fig. \ref{fig:mig_tech}, also defines allowed migration paths to achieve 100~Mbps deployment for residential users, business users and ITS MBSs.Some migrations, for example, migrating from FTTCab\_GPON\_25 to FTTH\_UDWDM\_100 or FTTB\_UDWDM\_50 to FTTB\_XGPON\_100 are restricted. This arises from our assumption that GPON and XGPON based technologies need a two-stage deployment (at least two remote nodes), to provide 100 Mbps to users, whereas UDWDM technologies can satisfy the same requirement using a single stage deployment \cite{patri2018,OpticalAccessSeamlessEvolution2011}. Migrations involving changes in deployment types would lead to a higher TCO and can be avoided by restricting migrations between similar deployment types.

In this paper, we evaluate two scenarios, both of which have to provide at least 100~Mbps per user~\cite{EuropeanCommissiononDigitalSingleMarket2014}:
\begin{enumerate}
    \item \textbf{Comparison scenario: fixed migration goal: FTTH}
    \\As most of the state-of-the-art papers consider a fixed final state as FTTH architectures, e.g., \cite{Reyes2014a} and \cite{VanderMerwe2009}, this scenario considers only FTTH-based PON technologies providing the end user with 100~Mbps.
    \item \textbf{Proposed scenario: flexible migration goal: any FTTx that maximizes the profitability}
    \\ All FTTx architectures, i.e., FTTCab, FTTB and FTTH, that provide the subscribers with 100~Mbps connection are considered.
\end{enumerate}

Comparing the results of these two scenarios, we can evaluate the influence of the final state, i.e., fixed (only FTTH) or flexible (any FTTx) final state, when it is chosen only based on the maximum NPV at the end of the network Life-cycle. As NPV requires a cost modelling of each of the different kinds of deployments, we need to find the TCO and define a revenue model, which is an input to the migration algorithm.

\subsection{Total Cost of Ownership and Revenue Model}
In order to make any business decisions, the cost of each option has to be evaluated. The TCO of the network consists of Capital and Operational Expenditures~(CAPEX and OPEX respectively). This paper is neither focused on the TCO calculation nor the cost modelling, but on the migration modelling and evaluation. However, as the TCO is crucial to select the best migration scenario, we introduce the used models, values and methodology. We closely follow the approach in \cite{VanderMerwe2009} and supplement it with values from \cite{CasierPhD2009, OpticalAccessSeamlessEvolution2011}. 

The CAPEX is split into two cost categories: civil works and equipment cost. The civil work cost is driven by the duct lengths and cost per meter of trenching and laying ducts. To find the costs of the network components, the respective electronic (Optical Line Terminal~(OLT) card, power splitter, Digital subscriber line access multiplexer~(DSLAM), ONU, etc.) and non-electronic (fiber lengths, cabinets, Heating Ventilation and Air Conditioning~(HVAC)) components are dimensioned based on the demands of the subscribers and the capacity of each component. The OPEX includes Energy, Fault Management (FM), Network Operation, Marketing and Rent. A detailed CAPEX and OPEX calculation used in this paper is presented in~\cite{patri2018}. It is important to note that throughout this work, we present all costs and revenues in Cost unit (C.U.), where 1 C.U. is fixed at the price of a single GPON ONU in the year 2013~\cite{OpticalAccessSeamlessEvolution2011}.  

The lengths of fibers and ducts for each of the deployments are obtained using the Automated Geography-Based Fixed Network Planning Tool~\cite{Grigoreva2016,elenagit2018}, which is based on the ESRI ArcMap~10.3.1\;\textregistered \cite{arcgis}. The tool calculates the duct and fiber lengths of feeder, distribution and last mile fibers, for the chosen area and network architecture. The planning is done on the real geographical street-based topologies to guarantee the meaningful results for the TCO calculations. The planning details are out of the scope of this paper but presented in detail in~\cite{patri2018}.

\begin{figure}[htbp!]
    \centering
    \captionsetup{justification=centering}
    \begin{tikzpicture}
\begin{axis}[
    scale only axis, 
    height=4cm,
    width=7 cm,
    legend style={legend pos=north west,font=\tiny},
	xlabel=Year,
	ylabel= \% Households connected,
	y tick label style={/pgf/number format/.cd,%
          scaled y ticks = false,
          set decimal separator={,},
          fixed},
      x tick label style={/pgf/number format/.cd,%
          scaled x ticks = false,
          set thousands separator={},
          fixed}]
\addplot[color=red,mark=x] coordinates {
	(2018,9600/29262)
	(2019,12800/29262)
(2020,17200/29262)
(2021,23300/29262)
(2022,31800/29262)
(2023,43500/29262)
(2024,59400/29262)
(2025,81300/29262)
(2026,110900/29262)
(2027,150900/29262)
(2028,204400/29262)
(2029,275100/29262)
(2030,367200/29262)
(2031,484900/29262)
(2032,631100/29262)
(2033,806800/29262)
(2034,1008700/29262)
(2035,1228700/29262)
(2036,1453500/29262)
(2037,1667400/29262)
(2038,1855900/29262)

};

\addplot[color=blue,mark=*] coordinates {
	(2018,9600/29262)
	(2019,13200/29262)
(2020,18300/29262)
(2021,25700/29262)
(2022,36300/29262)
(2023,51400/29262)
(2024,72700/29262)
(2025,102700/29262)
(2026,144600/29262)
(2027,202800/29262)
(2028,282200/29262)
(2029,388700/29262)
(2030,528100/29262)
(2031,704200/29262)
(2032,916600/29262)
(2033,1158000/29262)
(2034,1412600/29262)
(2035,1658200/29262)
(2036,1873000/29262)
(2037,2043100/29262)
(2038,2165800/29262)
};

\addplot[color=green,mark=o] coordinates {
	(2018,9600/29262)
	(2019,17900/29262)
(2020,34200/29262)
(2021,66100/29262)
(2022,128400/29262)
(2023,247300/29262)
(2024,466000/29262)
(2025,837300/29262)
(2026,1376000/29262)
(2027,1951700/29262)
(2028,2297300/29262)
(2029,2371300/29262)
(2030,2374700/29262)
(2031,2375300/29262)
(2032,2375600/29262)
(2033,2375800/29262)
(2034,2375900/29262)
(2035,2375900/29262)
(2036,2376000/29262)
(2037,2376000/29262)
(2038,2376000/29262)
};

\legend{Conservative,Likely, Aggressive}
\end{axis}
\end{tikzpicture}
    \caption{{Percentage of yearly connected residential subscribers 
    based on the penetration curves from~\cite{VanderWee2014}.} }
    \label{fig:pen_curves}
\end{figure}
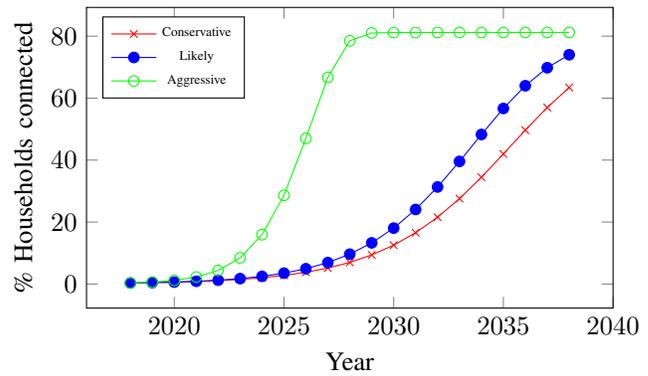

We model the subscriber behaviour with three different penetration curves, as proposed in~\cite{VanderWee2014}: "conservative", "realistic", and "aggressive", which differ on the subscriber joining rates. {These joining rates are then used to derive the number of subscribers connected to the network in every year, which is helpful in finding the yearly revenue generated in an area. Figure \ref{fig:pen_curves} shows the percentage of households connected in a pure Residential scenario at a given year. Readers may note that we limit the techno-economic calculations to 20 years, since projects in a given area last between 5-10 years and the remaining years are used to recover costs.}  

The Average Revenue Per User~(ARPU), shown in Table \ref{tab:revenue}, reflect the real optical access market tariffs~\cite{FTTHCouncilEurope2015}. For business and ITS subscribers, the yearly generated revenues are assumed to be higher since these subscribers need separate hardware and wavelengths for better security and faster fault reparation~\cite{FTTHCouncil2016a}. Since the subscriber pays for the data-rate, the ARPU does not change across different technologies offering the same data-rate.

Every year, apart from the number of subscribers joining the network, a certain number of the total connected subscribers leave it {due to marketing, costs, other requirements, etc. This variation in the total number of connected subscribers is called user churn $c$ and can change depending on the market, existing competitors, etc.} The user churn is defined as a Bernoulli process; the success probability deciding if the churn occurs. As the trials in each year are independent from each other, the overall process in the migration is memoryless and thus Markovian. 

\begin{table}[htbp!]
\small
\centering
\begin{tabular}{|c|c|c|c|}
\hline
\textbf{\begin{tabular}[c]{@{}c@{}}Data-rate \\ $[$Mbps$]$\end{tabular}} & \textbf{\begin{tabular}[c]{@{}c@{}}Residential\\ $[$C.U./year$]$\end{tabular}} & \textbf{\begin{tabular}[c]{@{}c@{}}Business\\ $[$C.U./year$]$\end{tabular}} & \textbf{\begin{tabular}[c]{@{}c@{}}ITS\\ $[$C.U./year$]$\end{tabular}} \\ \hline
20 Mbps & 3.6 & 3.6 & - \\ \hline
30 Mbps & 7.2 & 36 & - \\ \hline
50 Mbps & 10.8 & 84 & 84 \\ \hline
100 Mbps & 13.2 & 110 & 110 \\ \hline
\end{tabular}%
\caption{Yearly ARPU Summary, based on~\cite{FTTHCouncil2016a}.}
\label{tab:revenue}
\end{table}

The level of churn is assumed to be 10\%~\cite{OASE2013}. This user churn leads to lower revenue for the same operational expenditures, since the services or the connection to the subscriber exists, but there is no monetary benefit out of it. We do not consider the ITS MBSs as a part of churn, since we assume that once connected, the public ITS provider does not leave the network. With this assumption, the operator revenue in a single year $t$ is defined as:
\vspace{+1em}
\begin{equation}\label{eqn:yearly_rev}
    {R_{t}(\gamma_{t}) = \smashoperator{\sum_{q\in\;Sub}}(1-c)^{\gamma_t}n_{q,t}.R_{q}\;,}
\end{equation}

where $Sub\in\{Residential, Business\}$ and takes values from Table \ref{tab:revenue}, $n_{q,t}$ is the number of subscribers of type $q$ who are connected to the network at year $t$ {(calculated from the penetration curves shown in Fig.~\ref{fig:pen_curves})}, $R_{q}$ is the per subscriber yearly revenue of type $q$, $0\leq c\leq 1$ is the yearly churn rate of the subscribers and $\gamma_{t}$ is a binary variable, which is defined as follows:
\vspace{+2.0em}
\begin{equation}\label{eqn:bin_var_churn}
    \gamma_{t} = 
    \begin{cases}
    1\text{\;\;\;\;if\;churn occurs},  \\
    0 \text{\;\;\;\;otherwise}.
    \end{cases}
\end{equation}


\section{{Expectimax based Migration Algorithm}}\label{sec:mig_model}
This section introduces the proposed AI-based migration algorithm. Here we define the migration decision metric. Then, we introduce the modified Expectimax algorithm and apply it on a simple example. 

Like the Expectimax algorithm, we also used three different nodes (maximizer, chance and terminal), whose functions are briefly discussed in Section \ref{sec:rel_work}.
\vspace{-0.5em}

\subsection{Proposed Decision Metric and Utility Function}\label{subsec:utility}

In this work, we choose the Net Present Value~(NPV)~\cite{remer1995compendium} as our decision making metric. NPV is an economic metric, which shows the time value of money to valuate long-term projects, or project-profitability and it is defined as:

\begin{equation}\label{eqn:npv}
NPV = \smashoperator{\sum_{t=1}^{T_{NW}}}PV_t\;,
\end{equation}

where $t$ is the time iterator, $T_{NW}$ is the network life-cycle and $PV_t$ is the Present Value~(PV) of the net cash flows at the time $t$. PV is defined as follows:

\begin{equation}\label{eqn:pv}
    PV_t = \frac{R_t-I_t}{(1+d)^t}\;\;,
\end{equation}

where $t$ is the time iterator indicating the current time period, $R_t$ is the revenue at time $t$ (cash inflow) as defined by Eq.~\ref{eqn:yearly_rev}, $d$ is the discount rate of the project and $I_t$ is the investment made in time $t$ (cash outflow). This investment is the additional CAPEX required in order to migrate to a newer technology as well as the related OPEX costs for the deployed technology. We assume the discount rate to be fixed to 10\% \cite{Tahon2012,CasierPhD2009}.

In case migrations occur in an given year, we need to model the CAPEX part of the investments. For this, we define an additional fixed cost inherent to migrating from technology $s$ to technology $s'$ as: 
\begin{multline}\label{eqn:mig_capex}
M_{s,s'} = \kappa_{i,j}\Delta_{CW_{s,s'}}+ \upsilon_{p,q}\Delta_{Equip_{s,s'}} \\
\forall\;s,s'\in\{\text{Possible\_Technologies\}}, s\neq s'\;,
\end{multline}
{where $\Delta_{CW_{s,s'}}$ is the difference in CAPEX civil works when there is a change of architecture, provided by the binary variable $ \kappa_{i,j}$ defined in Eq.~\ref{eqn:kappa_civil}; and $\Delta_{Equip_{s,s'}}$ is the difference in CAPEX equipment when there is an upgrade of the delivered bandwidth, provided by the binary variable $\upsilon_{p,q}$ defined in Eq.~\ref{eqn:upsilon_elec}. These two CAPEX values $\Delta_{CW_{s,s'}}$ and $\Delta_{Equip_{s,s'}}$ follow the TCO cost modelling in \cite{VanderMerwe2009}}. 
These binary variables $\kappa_{i,j}$ and $\upsilon_{p,q}$ distinguish which network upgrades have to be conducted: civil work, equipment or both. 
\begin{equation}\label{eqn:kappa_civil}
    \kappa_{i,j} = 
    \begin{cases}
    1\text{\;\;\;\;if\;} i\smashoperator{\neq}j\;\forall\;i,j\in\{\text{FTTCab, FTTB, FTTH}\}  \\
    0 \text{\;\;\;\;otherwise}
    \end{cases}
\end{equation}
\vspace{+1.0em}
\begin{equation}\label{eqn:upsilon_elec}
    \upsilon_{p,q} = 
    \begin{cases}
    1\text{\;\;\;\;if\;} p\smashoperator{\neq}q\;\forall\;p,q\in\{\text{20, 25, 50, 100}\}\;Mbps  \\
    0 \text{\;\;\;\;otherwise}
    \end{cases}
\end{equation}

Based on the definitions above, {we define our \textbf{Utility Function $U$} for a current technology $s$ at year $t$ given a user churn $\gamma_t$,}
as follows:
\begin{multline}\label{eqn:new_eval_fn}
\small
    U(s,t,\gamma_t) = 
    \begin{cases}
    \text{\;}\smashoperator{\sum_{i=t}^{T_{NW}}}\frac{R_i(\gamma_i)-OPEX_i}{(1+d)^{T_{start}-i}}\\\text{\;\;\;\;\;\;\;\;\;\;\;\;\;\;\;\;\;\;\;\;\;\;if\;terminal node}\\
    \\
     \text{\;}\max\limits_{s'}\;[H(s',t+1)\\
     \text{\;\;\;\;}+\frac{R_t(\gamma_t)-OPEX_t}{(1+d)^{T_{start}-t}} -\mu_t\cdot M_{s,s'}]\\\text{\;\;\;\;\;\;\;\;\;\;\;\;\;\;\;\;\;\;\;\;\;\;if\;maximizer node},\\
    \end{cases}
\end{multline}
where $R_i(\gamma_i)$, $OPEX_i$ and $M_{s,s'}$ are the revenue, yearly OPEX and migration CAPEX respectively. $\gamma_i$ is a binary variable ($\gamma_i=1$ when user churn has occurred; $\gamma_i=0$ otherwise). $s$ is the current technology and $s'$ is the technology to be migrated to, $T_{start}$ is the starting year of the migration window, $d$ is the discount rate of the project, $\mu_t$ is a binary variable ($\mu_t=1$ when migration takes place in year $t$; $\mu_t=0$ otherwise). The network life-cycle is $T_{NW}$, which can be any period of years longer than $T_{mig}$. $H(s',t+1)$ is the expected present value generated by the child maximizer nodes at depth $t+1$ for the next technology $s'$.

At the maximizer nodes, the subtraction of $\mu_t\cdot M_{s,s'}$ implicitly refers to the capital expenditures required for migration. The value at chance nodes is the weighted average of the value of its children, given by Eq.~\ref{eqn:chance_node}.
\begin{equation}\label{eqn:chance_node}
    H(s,t) = \smashoperator{\sum_{\gamma_t~\in~\{0,1\} }}  Pr(\gamma_t)~{\cdot}~U(s,t,\gamma_t),
\end{equation}
where $H(s,t)$ is the value of the chance node of state $s$ at depth $t$. Looking at Eq.~\ref{eqn:new_eval_fn} and Eq.~\ref{eqn:chance_node}, it is evident that the algorithm is recursive in nature.

To model the uncertainty of churn, we define $Pr(\gamma_t)$ as a two state Markov Chain with steady-state probabilities as defined in Eq.~\ref{eqn:steady_state_prob}.
\begin{equation}\label{eqn:steady_state_prob}
Pr(\gamma_t)~=~
\begin{cases}
   0.9\text{\;\;\;\;if\;}\gamma_t~=~0\\
   0.1\text{\;\;\;\;if\;}\gamma_t~=~1\\
\end{cases}
\end{equation}

\subsection{Proposed Search Tree Algorithm}

To explain the algorithm, let us use a simple example of a migration from ADSL (20 Mbps) involving two technologies (see Fig.~\ref{fig:mig_tech}): FTTCab\_GPON\_25 (referred as PON1) and FTTB\_XGPON\_100 (referred as PON2). The migration window $T_{mig}$ is set to three years and the network life-cycle  $T_{NW}$ is set to five years. The final goal of the VIO is to find a technology, which provides it with the maximum profitability across the entire $T_{NW}$. Every year, for every possible migration, we have a maximizer node (blue rectangles with solid outlines in Fig. \ref{fig:build_tree}) and chance node (red rectangle with dashed outline). For subsequent years, there can be either terminal nodes (green rectangles with dotted outline) or maximizer nodes again. We first discuss how the tree is built and then move to its evaluation using the utility function defined in Subsection~\ref{subsec:utility}.

\subsubsection{Building the search tree}
\begin{figure}[htbp!]
  \begin{center}
    \centering
    \captionsetup{justification=centering}
    \includegraphics[width=6 cm,height=5.5 cm]{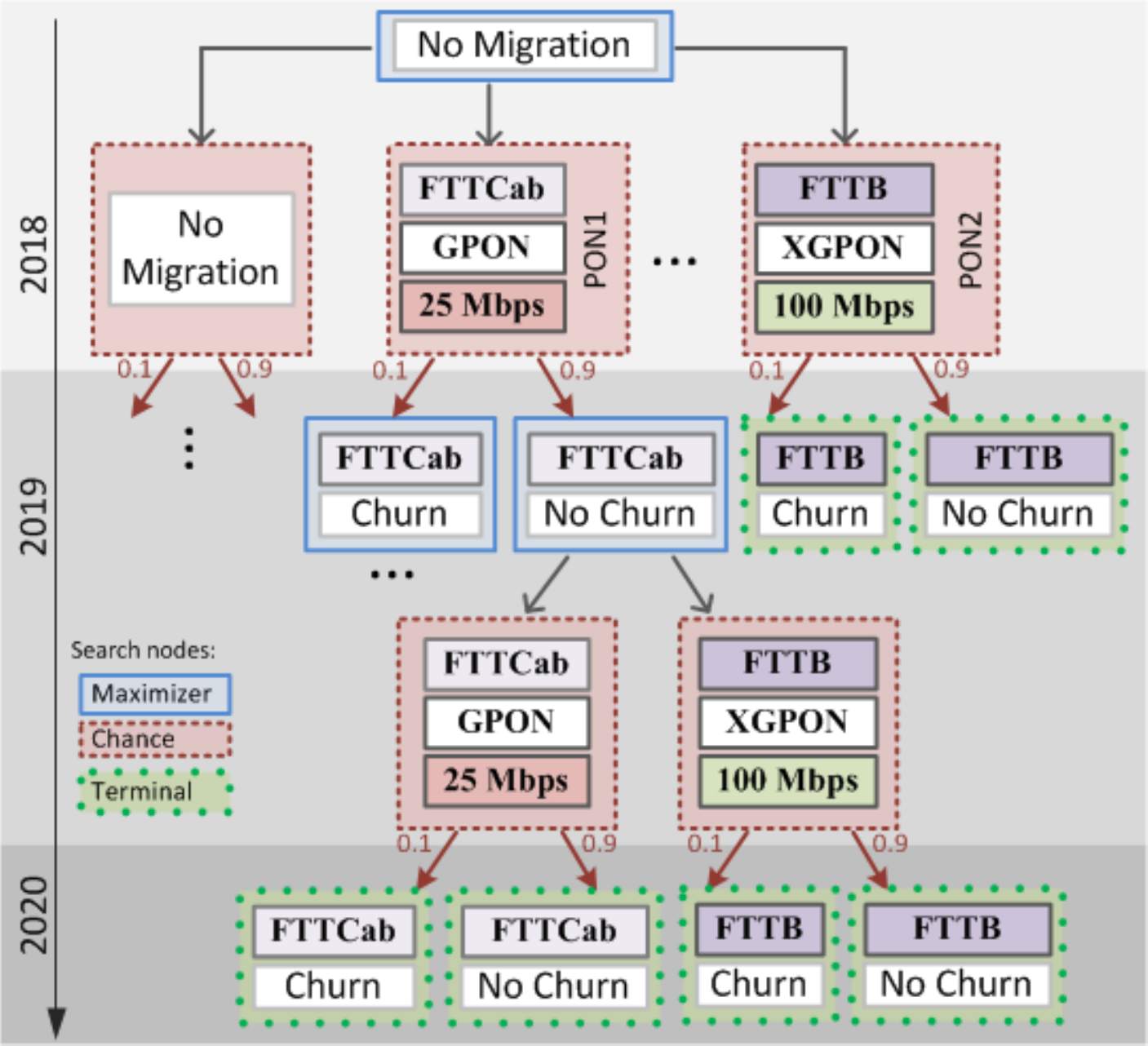}
    \caption{Top down building of search tree of depth three years and two technology choices. The tree is built recursively till a terminal condition is satisfied. }
    \label{fig:build_tree}
  \end{center}
\end{figure}

\begin{algorithm}\captionsetup{labelfont={sc,bf}, labelsep=newline}
    \scriptsize
    \caption{{Building the Expectimax tree}}
\begin{algorithmic}
\Ensure $curr\_year \leq T_{mig}$
\Procedure{add\_max}{$curr\_year, curr\_tech,type$}
    \For{$child$ in $possible\_migrations(curr\_tech)$}
        \State{\Call{add\_chance}{$child, curr\_year + 1$}}
    \EndFor
\EndProcedure
\State
\Procedure{add\_chance}{$tech,curr\_year$}
    \If{$tech.data\_rate$ is $100\;Mbps$}
        \State{\Call{add\_terminal}{$tech, curr\_year$}}
    
    \ElsIf{$curr\_year$ is $T\_mig$}
        \State{\Call{add\_terminal}{$tech, curr\_year$}}
    \Else
        \State{\Call{add\_max}{$curr\_year,curr\_tech,Churn$}}
        \State{\Call{add\_max}{$curr\_year,curr\_tech,No\_Churn$}}
    \EndIf
\EndProcedure
\State
\Procedure{add\_terminal}{$tech,curr\_year$}
    \State{\Call{begin\_eval}{$tech, curr\_year$}}
\EndProcedure

\end{algorithmic}
    \label{alg:build_tree}
\end{algorithm}

At the beginning of the migration period, a maximizer node is present and indicates the current state of migrations (No migrations). Each of its child nodes (chance nodes), is assigned a new technology, to which migration is possible. For each chance node, there are two possibilities, namely $Churn$ and $No\_Churn$, both having different utilities (refer Eq.~\ref{eqn:chance_node} and Eq.~\ref{eqn:steady_state_prob}). Fig. \ref{fig:build_tree} shows that PON2 Churn and No Churn in 2019 are terminal nodes instead of maximizer nodes. This happens because PON2 satisfies the goal of the VIO to provide 100~Mbps to its subscribers. 

Also, in 2019, PON1 has an option to migrate to either PON2 or not migrate at all. In the final year, all the technologies result in a terminal node, since the tree reaches the end of the migration window. {The flow of tree building is a recursive approach and is implemented according to the pseudo-code provided in Algorithm \ref{alg:build_tree}.}

\subsubsection{Evaluating the search tree}
\begin{figure}[htbp!]
  \begin{center}
    \centering
    \captionsetup{justification=centering}
    \includegraphics[width=6 cm,height=5.5 cm]{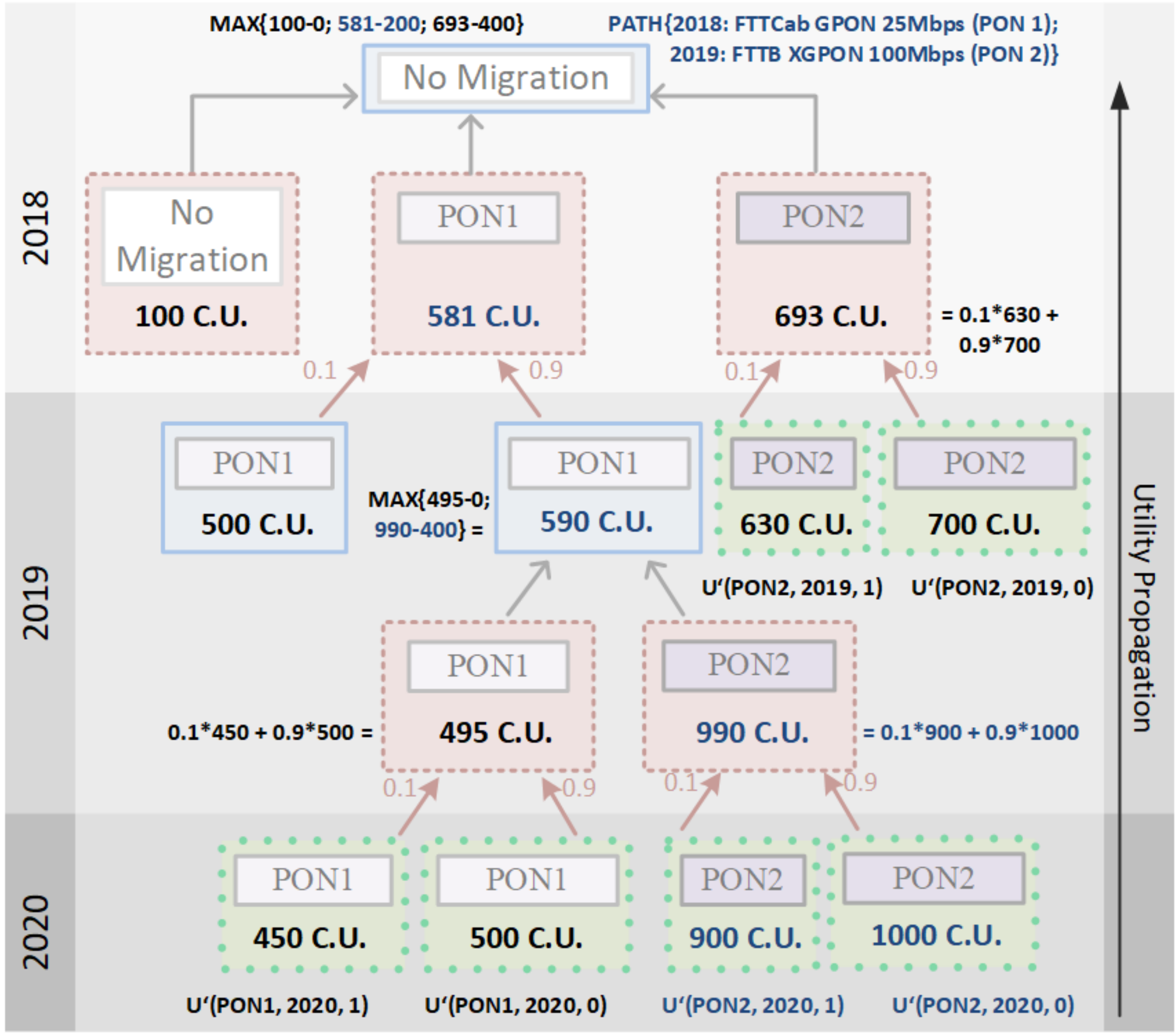}
    \caption{Bottom up evaluation of utility function in search tree reveals migration path (in blue), which provides the highest NPV at the end of the network life-cycle.}
    \label{fig:eval_tree}
  \end{center}
\end{figure}

After the tree is built, the evaluation finds the migration path with the highest expected utility value. For explaining this evaluation, we refer to Fig.~\ref{fig:eval_tree}. The algorithm always starts from the terminal nodes and then traverses its way up. The values stored at the terminal nodes are the NPV values from that year to end of $T_{NW}$ (using Eq.~\ref{eqn:new_eval_fn}), assuming no further churn occurs and the subscriber penetration follows the "realistic" curve. The chance nodes find the expected value of its children, which translates to the expected NPV from the current year to the end of the network life-cycle (using Eq.~\ref{eqn:chance_node}).

\begin{algorithm}\captionsetup{labelfont={sc,bf}, labelsep=newline}
    \caption{{Evaluating the Expectimax tree}}
    \scriptsize
\begin{algorithmic}
\Ensure $curr\_year \leq T_{mig}$
\Function{util\_fn}{$node\_type,year, tech,churn$}
    \If{$node\_type$ is $Terminal$}
        \State{\Return{$U(tech, year,churn)$}} 
        \State{\Comment{Ref. Equation \ref{eqn:new_eval_fn} with node type terminal}}
    
    \Else
        \State{$children = possible\_migrations(tech)$}
        \State{\Return{$\max\limits_{children}[U(child,year+1,churn)]$}} 
        \State{\Comment{Ref. Equation \ref{eqn:new_eval_fn} with node type maximizer}}
    \EndIf
\EndFunction
\end{algorithmic}
    \label{alg:eval_tree}
\end{algorithm}

At every maximizer node, migration costs are subtracted from the accumulated NPV, which can be seen from the PON1 maximizer node in year 2019. This is repeated till the the top-most maximizer node is reached, where the node with the maximum accumulated NPV is chosen. This leads to the search tree choosing the migration path as PON1 in 2019 and PON2 in 2020 as the most profitable decision, with an accumulated NPV of 381 C.U. {Algorithm \ref{alg:eval_tree} shows how a built tree is evaluated, based on the current node being evaluated.}

\section{Evaluation of the Proposed Migration Algorithm}\label{sec:results}

In our work, we have evaluated the performance of the proposed migration algorithm in a number of case studies. We looked into three types of deployment areas: dense urban~(New York), urban~(Munich), and suburban~(Ottobrunn)~\cite{Mitcsenkov2010}. 

Due to space limitations, we present the results of the Urban deployment scenario in Munich (Germany). All the other results have been showcased and thoroughly discussed in Chapter 6 of \cite{patri2018}. An interested reader can also reproduce these results using the source code {and input excel files (including migration matrices)} provided in \cite{saigit2018}. 

As already mentioned, three demand types have been considered: residential, business and public ITS base stations. The residential demands are defined by the buildings positions and the number of households per building, whereas the business demands are defined by the building positions and the density of the businesses in that area. For added security purposes, the business and ITS demands need to be sent on different wavelengths, which is supported by UDWDM and HPON technologies. 

In case of Munich, each building hosts a Multiple Dwelling Unit~(MDU), which connects an average of 6-8 potential residential or business subscribers~\cite{JLL2016, City}. These subscribers have been divided into residential and business subscribers using the fixed percentage of business subscribers in a city~\cite{rokkas2015techno, VanderMerwe2009}. In Munich, 7\% of the total buildings are considered to be business buildings~\cite{Tahon2012}. The considered demands for the converged migration are: 27213 residential, 2049 business and 2 ITS MBS.

All the presented results refer to the following scenario: area of $7~km^2$ in the center of Munich, migration starting in $T_{start}$=2018 and ending in 2027 (i.e., $T_{mig}$=10 years), $T_{NW}$ is set to 20 years, churn rate of $c$=10\% and churn probability $Pr(\gamma_t)$=0.1. The technological scenario in our case was migration from the existing copper, i.e., ADSL2+, network to a future-proof PON architecture~\cite{FTTHCouncilEurope2015,OASE2013}.

The migration goal for the residential and business subscribers is to offer at least 100~Mbps per household or business by 2025. This requirement was dictated by the EU Broadband Regulation Policy on Digital Single Market~\cite{EuropeanCommissiononDigitalSingleMarket2014}. The ITS Base stations, although not governed by any EU regulatory policy, have a technical requirement of at least 50~Mbps data-rate, to support current Public ITS demands \cite{Grigoreva2016}. However, a higher data-rate of 100~Mbps was assumed to be favourable in case of future increase in Public ITS traffic. 

\subsection{Pure Residential Migration}
Let us first consider a purely residential scenario, where all the demands in the network are residential subscribers, i.e., households, whose tariffs are provided in Table~\ref{tab:revenue}.

\begin{figure}[htbp!]

\centering
\captionsetup{justification=centering}
\begin{tikzpicture}
\pgfplotsset{every tick label/.append style={font=\small}}
\pgfplotstableread{ 
Technology	{Civil Works}	{Fiber Cost}	{Central Office}	{Remote Nodes}	Buildings
{FTTH\_UDWDM\_100} 	2.695375796	0.000499355	1.422214248	0.373180234	5.433668239
{FTTB\_UDWDM\_100} 	2.695375796	0.000499355	0.398617092	0.106622924	4.203403732
{FTTH\_XGPON\_100} 	5.000952574	0.000437926	0.683844804	0.273774862	5.221058027
{FTTB\_XGPON\_100} 	5.000952574	0.000437926	0.352766956	0.068443715	1.966666667
{FTTH\_HPON\_100} 	3.948139877	0.000353074	0.483904039	0.013156995	3.690793521
{FTTB\_HPON\_100} 	3.948139877	0.000325065	0.483904039	0.015942178	2.272571936
}\datatable

\begin{axis}[
    xbar stacked,   
    xmin=0,         
    ytick=data,     
    yticklabels from table={\datatable}{Technology},  
    xlabel={C.U. per subscriber passed},
    width=5.15 cm,
    height=5.5 cm,
    label style={font=\scriptsize},
    tick label style={font=\scriptsize},
    legend cell align=left,
    legend pos= outer north east,
    legend style={font=\scriptsize,draw=none}
]
\addplot [fill=blue!40] table [x={Civil Works}, y expr=\coordindex] {\datatable};    
\addlegendentry{Civil}
\addplot [fill=red!40]table [x={Fiber Cost}, y expr=\coordindex] {\datatable};
\addlegendentry{Fiber}
\addplot [fill=green!40] table [x={Central Office}, y expr=\coordindex] {\datatable};
\addlegendentry{CO}
\addplot [fill=magenta!80] table [x={Remote Nodes}, y expr=\coordindex] {\datatable};
\addlegendentry{RN}
\addplot [fill=yellow!40] table [x=Buildings, y expr=\coordindex] {\datatable};
\addlegendentry{Buildings}

\end{axis}
\end{tikzpicture}
\caption{CAPEX categories of selected deployments in Munich pure residential scenario offering 100 Mbps to subscribers.}
\label{fig:capex_munich_res}
\end{figure}
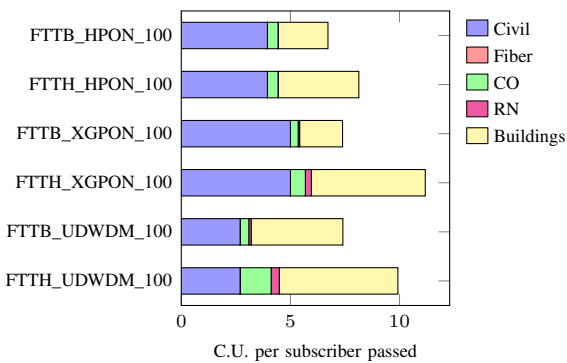

Out of the 13 different PON technologies which can be derived from the migration tree shown in Fig. \ref{fig:mig_tech}, Fig. \ref{fig:capex_munich_res} shows the per subscriber CAPEX in C.U. of selected technologies, all of which offer at least 100 Mbps to the subscriber. {The CAPEX costs are divided into components like Civil Works, Fiber laying costs and equipment costs at various locations like Central Office (CO), Remote Nodes (RN) and buildings.} We observed that FTTH technologies are more expensive than non-FTTH ones, with FTTH\_XGPON\_100 being the most expensive technology to deploy (11.18 C.U. per subscriber). {Due to a single stage deployment and a high subscriber density in urban areas, UDWDM based FTTB/FTTH technologies have lower civil works cost as compared to XGPON based FTTB/FTTH technologies.}

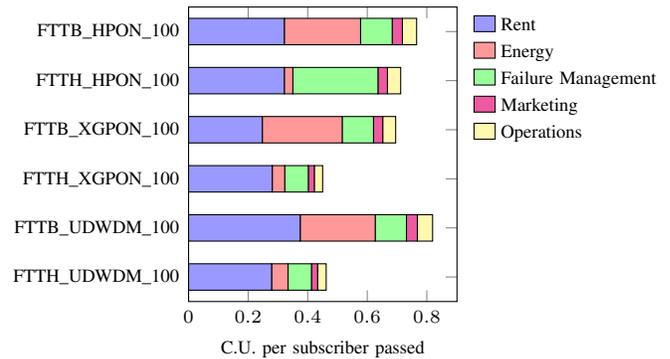
\begin{figure}[htbp!]
\centering
\captionsetup{justification=centering}
\begin{tikzpicture}
\pgfplotsset{every tick label/.append style={font=\small}}
\pgfplotstableread{ 
Technology	Rent	Energy	FM	Marketing	Operations
{FTTH\_UDWDM\_100}	0.278783405	0.055152974	0.078644922	0.020629065	0.028880691
{FTTB\_UDWDM\_100}	0.37485066	0.251807341	0.104933924	0.036579596	0.051211435
{FTTH\_XGPON\_100}	0.280956872	0.041872692	0.07915256	0.020099106	0.028138749
{FTTB\_XGPON\_100}	0.247557925	0.268193947	0.105191198	0.031047153	0.043466015
{FTTH\_HPON\_100}	0.321419589	0.028738979	0.285786154	0.031797236	0.04451613
{FTTB\_HPON\_100}	0.321419589	0.25604634	0.106279663	0.03418728	0.047862191
}\datatable

\begin{axis}[
    xbar stacked,   
    xmin=0,         
    ytick=data,     
    yticklabels from table={\datatable}{Technology},  
    xlabel={C.U. per subscriber passed},
    width=5.15 cm,
    height=5.5 cm,
    label style={font=\scriptsize},
    tick label style={font=\scriptsize},
    legend cell align=left,
    legend pos= outer north east,
    legend style={font=\scriptsize,draw=none}
]
\addplot [fill=blue!40] table [x=Rent, y expr=\coordindex] {\datatable};    
\addlegendentry{Rent}
\addplot [fill=red!40]table [x=Energy, y expr=\coordindex] {\datatable};
\addlegendentry{Energy}
\addplot [fill=green!40] table [x=FM, y expr=\coordindex] {\datatable};
\addlegendentry{Failure Management}
\addplot [fill=magenta!80] table [x=Marketing, y expr=\coordindex] {\datatable};
\addlegendentry{Marketing}
\addplot [fill=yellow!40] table [x=Operations, y expr=\coordindex] {\datatable};
\addlegendentry{Operations}

\end{axis}
\end{tikzpicture}
\caption{OPEX categories of selected deployments in Munich pure residential scenario offering 100 Mbps to subscribers in year 2018}
\label{fig:opex_munich_res}
\end{figure}

From Fig. \ref{fig:opex_munich_res}, we see that FTTH technologies have the least OPEX, around 0.6 C.U. per subscriber. This is because the subscribers pay for the rent and energy costs of the ONU. The highest OPEX of about 1 C.U. per subscriber per year is from FTTB\_UDWDM\_100, since the energy and rent costs of expensive UDWDM ONUs are borne by the VIO. To supplement the migration algorithm, we fix the per subscriber OPEX in ADSL (copper) to 0.25 C.U. The OPEX calculations in our work are based on the model and data provided in \cite{OpticalAccessSeamlessEvolution2011}. The results of the expected NPV and the migration path for each of the different subscriber penetration curve is shown in Table~\ref{tab:muc_res_result}.

\begin{table}[htbp!]
\tiny
\centering
\captionsetup{justification=centering}
\resizebox{\columnwidth}{!}{%
\begin{tabular}{|c|c|c|}
\hline
\multicolumn{3}{|l|}{\textbf{FTTCab/FTTB/FTTH provide 100 Mbps}}                                                                                               \\ \hline
\textbf{Penetration Curve} & \textbf{FTTx Migration Path}                                                           & \textbf{Net Present Value {[}C.U.{]}}           \\ \hline
Conservative               & \begin{tabular}[c]{@{}c@{}}2019: FTTB\_UDWDM\_50\\ 2020: FTTB\_UDWDM\_100\end{tabular} & {\textbf{93837}}  \\ \hline
Realistic                     & \begin{tabular}[c]{@{}c@{}}2019: FTTB\_UDWDM\_50\\ 2020: FTTB\_UDWDM\_100\end{tabular} & {\textbf{180161}} \\ \hline
Aggressive                 & 2019: FTTB\_UDWDM\_100                                                                 & {\textbf{886778}} \\ \hline
\multicolumn{3}{|l|}{\textbf{Only FTTH provide 100 Mbps}}                                                                                                    \\ \hline
\textbf{Penetration Curve} & \textbf{{FTTH} Migration Path}                                                           & \textbf{Net Present Value {[}C.U.{]}}          \\ \hline
Conservative               & 2019: FTTH\_HPON\_100                                                              & {\textbf{13313}}  \\ \hline
Realistic                     & 2019: FTTH\_HPON\_100                                                              & {\textbf{100679}} \\ \hline
Aggressive                 & 2019: FTTH\_HPON\_100                                                              & { \textbf{815178}} \\ \hline
\end{tabular}%
}
\caption{Technologies deployed in each scenario for Munich pure residential scenario.}
\label{tab:muc_res_result}
\end{table}

From the results in Table \ref{tab:muc_res_result}, we observe that due to low ARPU for residential subscribers, the VIO cannot have a higher NPV in a purely residential deployment, unless the subscriber penetration is aggressive. We see that when only FTTH technologies provide 100 Mbps, the NPV is 20-50\% lower because non-FTTH technologies with 100 Mbps (like FTTB\_UDWDM\_100) have a higher return on investment. We also observe that the algorithm suggests migrations in the beginning of the migration window, in order to get more revenue from the subscribers. This behaviour is also recorded by the heuristic optimization models of \cite{Turk2013}.

Overall, in the worst case of a conservative subscriber penetration rate, coupled with only FTTH technologies providing 100 Mbps data-rates, early migrations to a hybrid PON optical architecture result in at least a minimally positive NPV.

\subsection{Converged Migration}
\begin{figure}[htbp!]
  \begin{center}
    \centering
    \captionsetup{justification=centering}
    \includegraphics[width=7cm,height=4.5cm]{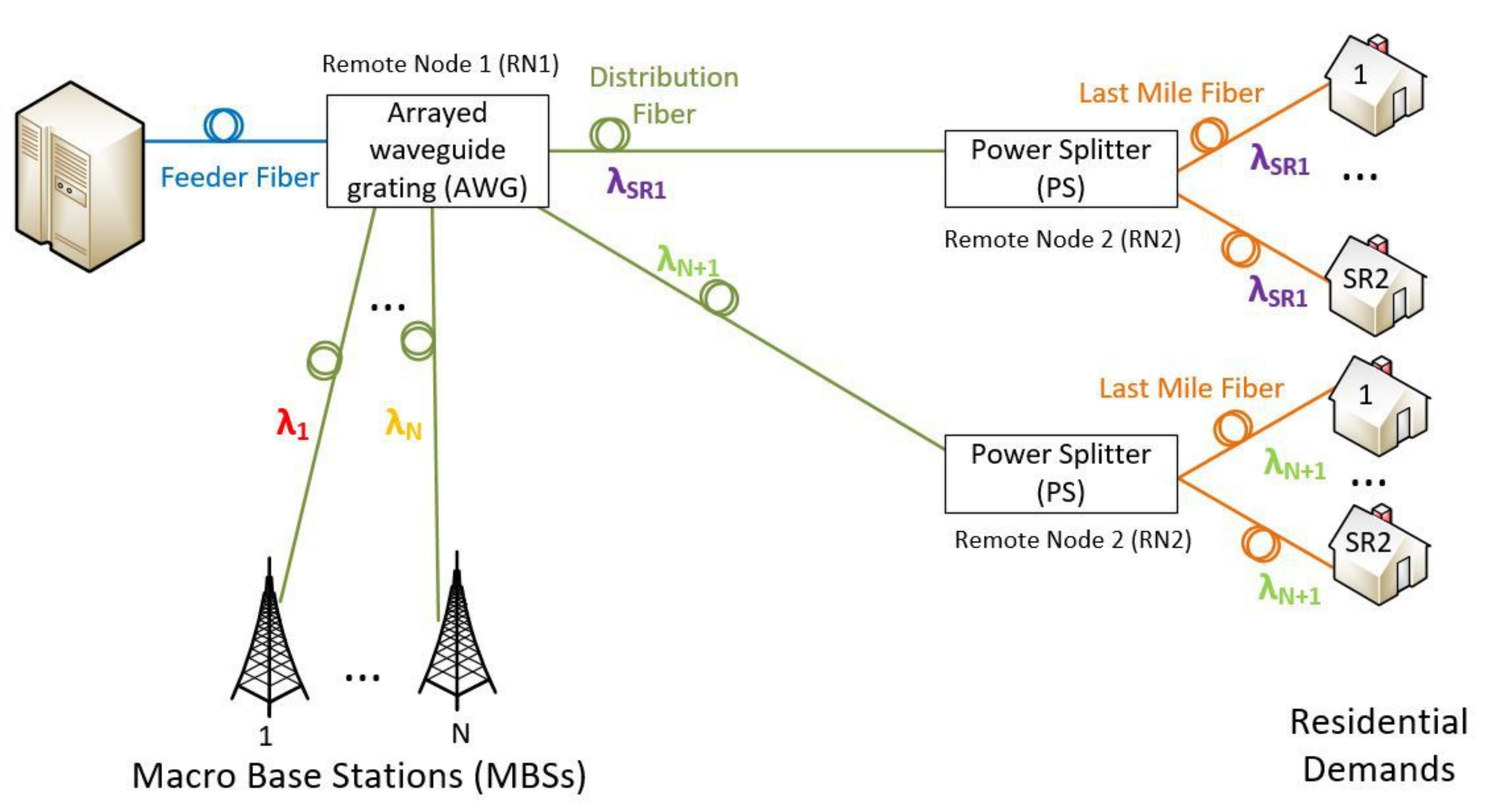}
    \caption{Example deployment of a two-stage FTTH\_HPON\_100 with a 1:80 AWG at RN1 and 1:16 Power Splitters at RN2 in a converged scenario.}
    \label{fig:ftth_hpon}
  \end{center}
\end{figure}

Let us now consider a converged scenario all different demands: residential, business as well as public ITS MBSs have to be satisfied. First, the planning and cost evaluation are performed and the results show that both CAPEX and OPEX for every architecture in the converged scenario, are between 5-10\% higher than in the pure residential scenario. While deploying the network, it is made sure that the business subscribers and residential subscriber do not share equipment for security reasons~\cite{FTTHCouncil2016a}.
\begin{table}[htbp!]
\tiny
\captionsetup{justification=centering}
\centering
\resizebox{\columnwidth}{!}{%
\begin{tabular}{|c|c|c|}
\hline
\multicolumn{3}{|l|}{\textbf{FTTCab/FTTB/FTTH provide 100 Mbps}}                                                                                                        \\ \hline
\textbf{Penetration Curve} & \textbf{FTTx Migration Path}                                                                   & \textbf{Net Present Value {[}C.U.{]}}   \\ \hline
Conservative               & \begin{tabular}[c]{@{}c@{}}2019: FTTB\_HPON\_50\\ 2020: FTTB\_HPON\_100\end{tabular} & {\textbf{190069}}  \\ \hline
Realistic                     & \begin{tabular}[c]{@{}c@{}}2019: FTTB\_HPON\_50\\ 2020: FTTB\_HPON\_100\end{tabular} & {\textbf{325132}}  \\ \hline
Aggressive                 & 2019: FTTB\_UDWDM\_100                                                                     & { \textbf{1429279}} \\ \hline
\multicolumn{3}{|l|}{\textbf{Only FTTH provide 100 Mbps}}                                                                                                             \\ \hline
\textbf{Penetration Curve} & \textbf{{FTTH} Migration Path}                                                                   & \textbf{Net Present Value {[}C.U.{]}}   \\ \hline
Conservative               & 2019: FTTH\_UDWDM\_100                                                                          & { \textbf{99885}}   \\ \hline
Realistic                     & 2019: FTTH\_UDWDM\_100                                                                          & { \textbf{236452}}  \\ \hline
Aggressive                 & 2019: FTTH\_UDWDM\_100                                                                          & {\textbf{1352004}} \\ \hline
\end{tabular}%
}
\caption{Technologies deployed in each migration scenario for Munich converged scenario.}
\label{tab:muc_conv_result}
\end{table}
To visualize better a converged network deployment, let us focus on the example of FTTH\_HPON\_100, which has been depicted in Fig.~\ref{fig:ftth_hpon}. To meet the demands, every 10~Gbps card at the OLT is connected to a single Arrayed Waveguide Grating (AWG) at the first RN. This device splits the optical signal from feeder fiber to up to 80~different wavelengths which can be either split further into 100~Mbps connections at the second RNs or can be connected directly to ITS LTE MBSs.

In this deployment, additional infrastructure and equipment was added in order that each business building (consisting of 6-8 business subscribers) gets its own wavelength, in accordance with the Service Level agreements (SLAs) that justify the higher price~\cite{FTTHCouncil2016a}. Due to these factors, the NPV is on an average 1.6 to 7.5 times higher for the converged case in all different scenarios, as compared to a pure residential scenario. The results from the migration algorithm are shown in Table \ref{tab:muc_conv_result}.

In case migrations to FTTx technologies are allowed, FTTB\_HPON\_100 is preferred due to its highest NPV. Compared to the pure residential scenario, the algorithm selects HPON for "conservative" penetration curves, since they are more cost effective as compared to GPON and UDWDM technologies. Among FTTH technologies, migration to FTTH\_UDWDM\_100 is preferred because the high operational costs are offset by charging a higher tariff from business and ITS subscribers. FTTH\_HPON\_100, which was the preferred technology in the residential scenario, loses out marginally to FTTH\_UDWDM\_100 in this case, due to higher ONU costs in the hybrid PON technology.

\subsection{Sensitivity Analysis}
The sensitivity analysis has been applied to the converged planning scenario of Munich (Urban) area. The other scenarios have been also implemented, tested and found to behave in the same way as the one presented in this section. Here we chose a positive scenario, where there are no constraints on when or to which technology to migrate. Since different studies have different units of currency, like Pounds, Euros and Cost Unit, we converted all the currencies into cost units using the current currency conversion. Here 1 C.U. is fixed at 50 Euros and 44.97 GBP.

\renewcommand{\arraystretch}{1.1}
\begin{table}[htbp!]
\centering
\resizebox{\columnwidth}{!}{%
\begin{threeparttable}
\begin{tabular}{|c|c|c|c|c|}
\hline
\textbf{Component} & \textbf{OASE [C.U.]} & \textbf{Phillipson [C.U.]} & \textbf{Rokkas [C.U.]} & \textbf{BSG [C.U.]} \\ \hline
Fiber Duct & \normalsize{1.12 /m} & \normalsize{0.54 /m} & \normalsize{0.7 /m} & \normalsize{1.42} /m \\[5pt] \hline
Fiber & \normalsize{0.02 /m} & \normalsize{0.006 /m} & \normalsize{0.006 /m} & \normalsize{0.192 /m}\\[5pt] \hline
GPON OLT Card & \normalsize{40} & \normalsize{50} & \normalsize{70} & \normalsize{288} \\[5pt] \hline
XGPON OLT Card & \normalsize{80} & \normalsize{55}\tnote{*} & \normalsize{200} & \normalsize{300}\tnote{*} \\[5pt] \hline
WDM OLT Port Card & \normalsize{8.8} & \normalsize{60}\tnote{*} & \normalsize{200} & \normalsize{350}\tnote{*} \\[5pt] \hline
Power Splitter & \normalsize{1.8} & \normalsize{2}\tnote{*} & \normalsize{10} & \normalsize{1.4}\tnote{*} \\[5pt] \hline
AWG & \normalsize{2.2} & \normalsize{2}\tnote{*} & \normalsize{12}\tnote{*} & \normalsize{2}\tnote{*} \\[5pt] \hline
DSLAM+Cabinet & \normalsize{124} & \normalsize{220} & \normalsize{300} & \normalsize{294} \\[5pt] \hline
GPON ONU & \normalsize{1} & \normalsize{5} & \normalsize{2} & \normalsize{1.6} \\[5pt] \hline
XGPON ONU & \normalsize{1.8} & \normalsize{5}\tnote{*} & \normalsize{4} &\normalsize{ 1.8}\tnote{*} \\[5pt] \hline
WDMPON ONU & \normalsize{2.3} & \normalsize{5}\tnote{*} & \normalsize{5}\tnote{*} & \normalsize{2.3}\tnote{*} \\[5pt] \hline
HPON ONU & \normalsize{3.1} & \normalsize{5.5}\tnote{*} & \normalsize{5} & \normalsize{3.1}\tnote{*} \\[5pt] \hline
\end{tabular}%
\begin{tablenotes}
  \item[*] Assumed according to model trends.
 \end{tablenotes}
 \end{threeparttable}
}
\caption{Per Unit Component Costs from OASE~\cite{OASE2013}, Phillipson~\cite{phillipson2013fourth} Rokkas~\cite{rokkas2015techno}, BSG~\cite{AnalysysMason2008}.}
\label{tab:comp_cost_vals}
\end{table}
\subsubsection{Component Costs}
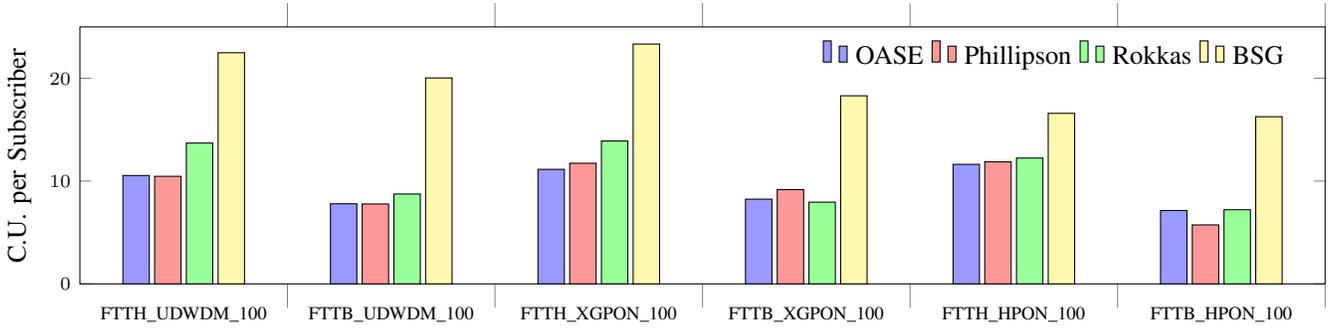
\begin{figure*}[htbp!]
\pgfplotstableread{
0	10.54204918	10.46278276	13.697269	22.48767615
1	7.79042933	7.771715851	8.742993833	20.03616907
2	11.14275616	11.73731371	13.8996779	23.33404265
3	8.241167069	9.168965683	7.94916187	18.30178238
4	11.62945519	11.87101078	12.23929349	16.59155403
5	7.138114882	5.736204034	7.211441176	16.26313592
}\dataset
\begin{tikzpicture}
\pgfplotsset{every tick label/.append style={font=\scriptsize}}

\begin{axis}[ybar,
        width=\textwidth,
        height=5.0cm,
        ymin=0,
        ymax=25,        
        ylabel={C.U. per Subscriber},
        xtick=data,
        xticklabels = {
            \strut {FTTH\_UDWDM\_100},
            \strut {FTTB\_UDWDM\_100},
            \strut {FTTH\_XGPON\_100},
            \strut {FTTB\_XGPON\_100},
            \strut {FTTH\_HPON\_100},
            \strut {FTTB\_HPON\_100}
        },
        major x tick style = {opacity=0},
        minor x tick num = 1,
        minor tick length=2ex,
        every node near coord/.append style={
                anchor=west,
                rotate=90
        },
        legend entries={OASE, Phillipson, Rokkas, BSG},
        legend columns=4,
        legend style={draw=none,nodes={inner sep=3pt}},
        ]
\addplot[draw=black,fill=blue!40] table[x index=0,y index=1] \dataset; 
\addplot[draw=black,fill=red!40] table[x index=0,y index=2] \dataset; 
\addplot[draw=black,fill=green!40] table[x index=0,y index=3] \dataset; 
\addplot[draw=black,fill=yellow!40] table[x index=0,y index=4] \dataset;
\end{axis}
\end{tikzpicture}
\captionsetup{justification=centering}
\caption{CAPEX of selected deployments for different component costs in a converged scenario (OASE~\cite{OASE2013}, Phillipson~\cite{phillipson2013fourth}, Rokkas~\cite{rokkas2015techno}, BSG~\cite{AnalysysMason2008}).}\label{fig:sense_analy_cost_per_hh}

\end{figure*}

Since techno-economic studies are specific to the costs involved in an area, different cost models have different expected NPVs. In this study, we took the costs mentioned in three different studies, namely Rokkas~\cite{rokkas2015techno} (deploying GPON based technology in a generic urban area), Phillipson~\cite{phillipson2013fourth} (Deploying HPON based technology in a Dutch city) and BSG~\cite{AnalysysMason2008} (FTTCab based deployment in London). As a further contribution, we collected data from each of these different studies and undertook the cost modelling for all the different deployment technologies. We then compared all the technologies with each other and with also the base case, which are the values taken from \cite{OpticalAccessSeamlessEvolution2011}. 

For each of the four studies, we run the migration algorithm and find the expected NPV and the migration steps. Both \cite{rokkas2015techno} and \cite{phillipson2013fourth} are comparable in terms of cost to the OASE base model \cite{OASE2013}. Hence the results can vary according to the costs. Table \ref{tab:comp_cost_vals} gives a list of major component costs from each of the studies. Since all components used in our work were not mentioned in every study, we assumed these components to follow model trends. Fig. \ref{fig:sense_analy_cost_per_hh} shows the per subscriber CAPEX in cost units for selected 100 Mbps deployments in a converged scenario. It is clear that three of the four studies have comparable costs, with the only exception being BSG.

We see that as the component costs increases, the NPV decreases, which is the expected behaviour. As seen in Table \ref{tab:comp_cost_result}, except the OASE base case, the other components choose FTTB\_HPON\_100 as the final technology and prefer early migrations to reap maximum benefits. The only exception to this is the BSG costs in conservative and realistic scenarios, which suggests not undertaking any migrations. This is because of the high costs involved and no change in the ARPU. Since the BSG study was done for an expensive and densely populated city like London, it is possible that the ARPU would be higher than what is used in our analysis.

\begin{table}[htbp!]
\tiny
\centering
\resizebox{\columnwidth}{!}{%
\begin{tabular}{|c|c|c|}
\hline
\multicolumn{3}{|l|}{\textbf{OASE}\cite{OASE2013}}                                                                                                                                   \\ \hline
\textbf{Penetration Curve} & \textbf{FTTx Migration Path}                                                                   & \textbf{Net Present Value {[}C.U.{]}}   \\ \hline
Conservative               & \begin{tabular}[c]{@{}c@{}}2019: FTTB\_HPON\_50\\ 2020: FTTB\_HPON\_100\end{tabular} & \textbf{190069}  \\ \hline
Realistic                     & \begin{tabular}[c]{@{}c@{}}2019: FTTB\_HPON\_50\\ 2020: FTTB\_HPON\_100\end{tabular} & \textbf{325132}  \\ \hline
Aggressive                 & 2019: FTTB\_UDWDM\_100                                                                         & {\color[HTML]{333333} \textbf{1429279}} \\ \hline
\multicolumn{3}{|l|}{\textbf{Phillipson}\cite{phillipson2013fourth}}                                                                                                                             \\ \hline
\textbf{Penetration Curve} & \textbf{FTTx Migration Path}                                                                   & \textbf{Net Present Value {[}C.U.{]}}   \\ \hline
Conservative               & 2019:FTTB\_HPON\_100           & \textbf{122363}  \\ \hline
Realistic                     & 2019:FTTB\_HPON\_100           & \textbf{221742}  \\ \hline
Aggressive                 & 2019:FTTB\_HPON\_100           & \textbf{1055627} \\ \hline
\multicolumn{3}{|l|}{\textbf{Rokkas}\cite{rokkas2015techno}}                                                                                                                                 \\ \hline
\textbf{Penetration Curve} & \textbf{FTTx Migration Path}                                                                   & \textbf{Net Present Value {[}C.U.{]}}   \\ \hline
Conservative               & 2019:FTTB\_HPON\_100 & \textbf{75805}                         \\ \hline
Realistic                     & 2019:FTTB\_HPON\_100 & \textbf{175184}                         \\ \hline
Aggressive                 & 2019:FTTB\_HPON\_100 & \textbf{1009069}                        \\ \hline
\multicolumn{3}{|l|}{\textbf{BSG}~\cite{AnalysysMason2008}}                                                                                                                                    \\ \hline
\textbf{Penetration Curve} & \textbf{FTTx Migration Path}                                                                   & \textbf{Net Present Value {[}C.U.{]}}   \\ \hline
Conservative               & No Migrations                                                                                  & \textbf{37421}                          \\ \hline
Realistic                     & No Migrations                                                                     & \textbf{53510}                          \\ \hline
Aggressive                 & 2019: FTTB\_HPON\_100                                                                           & \textbf{769030}                        \\ \hline
\end{tabular}%
}
\caption{Migration results of different component costs.}
\label{tab:comp_cost_result}
\end{table}

\subsubsection{OPEX}

The final part of the sensitivity analysis includes an OPEX study. In many techno-economic works including~\cite{VanderMerwe2009} and~\cite{Tahon2012}, OPEX is considered as a fractional quantity of CAPEX, which was originally modelled in \cite{verbrugge2006methodology}. This is because techno-economic researchers do not have access to component specific data like mean time to repair, energy consumption, component footprint and technician salaries.

Here, we choose two different OPEX models. The base model is already defined in Section \ref{sec:methodology}. The percentage based OPEX model is taken from \cite{Tahon2012}, where the OPEX of a technology $t$ is defined as follows.
\begin{equation}\label{eqn:new_opex}
    OPEX_t = 0.1*C_{Elec_t} + 0.01*C_{CW_t}
\end{equation}
where $C_{Elec_t}$ and $C_{CW_t}$ are the electronic and the civil works CAPEX of a technology $t$, respectively.

\begin{table}[htbp!]
\tiny
\resizebox{\columnwidth}{!}{%
\begin{tabular}{|c|c|c|}
\hline
\multicolumn{3}{|l|}{\textbf{Base OPEX Model}}                                                                                                                 \\ \hline
\textbf{Penetration Curve} & \textbf{FTTx Migration Path}                                                         & \textbf{Net Present Value {[}C.U.{]}}      \\ \hline
Conservative               & \begin{tabular}[c]{@{}c@{}}2019: FTTB\_HPON\_50\\ 2020: FTTB\_HPON\_100\end{tabular} & {\color[HTML]{000000} \textbf{152401.51}}  \\ \hline
Realistic                     & \begin{tabular}[c]{@{}c@{}}2019: FTTB\_HPON\_50\\ 2020: FTTB\_HPON\_100\end{tabular} & {\color[HTML]{000000} \textbf{272726.55}}  \\ \hline
Aggressive                 & 2019: FTTH\_UDWDM\_100                                                               & {\color[HTML]{000000} \textbf{1282631.54}} \\ \hline
\multicolumn{3}{|l|}{\textbf{Percentage OPEX Model}}                                                                                                           \\ \hline
\textbf{Penetration Curve} & \textbf{FTTx Migration Path}                                                         & \textbf{Net Present Value {[}C.U.{]}}      \\ \hline
Conservative               & \begin{tabular}[c]{@{}c@{}}2019: FTTB\_HPON\_50\\ 2020: FTTB\_HPON\_100\end{tabular} & {\color[HTML]{000000} \textbf{216007.47}}  \\ \hline
Realistic                     & \begin{tabular}[c]{@{}c@{}}2019: FTTB\_HPON\_50\\ 2020: FTTB\_HPON\_100\end{tabular} & {\color[HTML]{000000} \textbf{359372.65}}  \\ \hline
Aggressive                 & 2019:FTTH\_UDWDM\_100                                                                & {\color[HTML]{000000} \textbf{1535318.70}} \\ \hline
\end{tabular}%
}
\caption{Migration results for different OPEX models.}
\label{tab_mig_result_opex}
\end{table}

As seen in Eq.~\ref{eqn:new_opex}, the OPEX is directly proportional to the CAPEX values shown in Fig. \ref{fig:capex_munich_res}. For most technologies, the percentage based OPEX is cheaper per subscriber connected. However, in the case of FTTH based technologies, the newer OPEX model is between 0.25-1 C.U. higher, for every subscriber connected. Hence, the benefits of OPEX savings due to lower energy costs in FTTH architectures is not considered in this fraction based model. 

From Table \ref{tab_mig_result_opex} we can infer that for a rough analysis, a percentage based OPEX model could be considered, keeping in mind the risk of underestimating the cost factors of various technologies.

\section{Conclusions and Outlook}\label{sec:conclusions}
Network operators have to timely upgrade their access network infrastructure in order to provide the required services and stay profitable. This motivation defines our approach to the complex strategic multi-period migration problem. We put the primary goal of the operator, profitability in terms of maximizing the Net Present Value, as our utility function for the migration algorithm. For the adequate calculation of the utility, we calculate the realistic deployment and operations costs, take into account subscriber penetration, subscriber churn and realistic revenue. 

{In this paper, we have proposed a migration algorithm based on a modified Expectimax search, to find the migration paths.} We show that the proposed flexible final state results in up to 50\% profitability increase over a typical state-of-the-art fixed state migrations. We also avoid overestimating the revenue by including the user churn in our model. Finally, we have validated our assumptions with the sensitivity analysis.

{The proposed migration algorithm is flexible and can be used for a migration study with any user defined migration window}. It could be improved with a heuristic-based pruning of the search tree in order to reduce time complexity. For a future upgrade, the churn probability can be modelled using a more realistic model. Real world access network migration data may be used to get more accurate estimates or to develop a generalized heuristic model. {Further, percentage based mixed cases could be added to the model as an input and the results of these partial migration scenarios evaluated.} The revenue model can be also be further investigated for optimal results. Finally, reliable predictions can be made using machine learning algorithms like random forest, if sufficient migration data is present. 

\ifCLASSOPTIONcaptionsoff
  \newpage
\fi



\bibliographystyle{IEEEtran}
\bibliography{Bibliography}

\end{document}